\documentclass[twocolumn,twoside,slac_two]{revtex4}
\usepackage{graphicx}
\usepackage{fancyhdr}
\begin{document}

\title{From E.\ Fermi to Fermi-LAT:\\ watching particle acceleration in supernova remnants}

%

\author{D. Caprioli}
\affiliation{Princeton University, 4 Ivy Ln., Princeton, NJ, 08544, USA}

\begin{abstract}
Supernova remnants (SNRs) have been regarded for many decades as the sources of Galactic cosmic rays (CRs) up to a few PeV. 
However, only with the advent of Fermi-LAT it has been possible to detect  --- at least in some SNRs --- $\gamma$-rays whose origin is unequivocally hadronic, namely due to the decay of neutral pions produced by collisions between relativistic nuclei and the background plasma. 
When coupled with observations in other bands (from radio to TeV $\gamma$-rays), Fermi-LAT data  present evidence for CR spectra significantly steeper than the standard prediction of  diffusive shock acceleration, forcing us to rethink our theoretical understanding of efficient particle energization at strong shocks. 
We outline how, by including the effects of CR-triggered magnetic field amplification, it is possible to reconcile non-linear models of diffusive shock acceleration with $\gamma$-ray observations, in particular providing a successful application of such a theory to Tycho's SNR.
Finally, we show how kinetic simulations can investigate the microphysics of the non-linear coupling of accelerated particles and magnetic fields, probing from first principles the efficiency of the Fermi mechanism at strong shocks 
\end{abstract}

\maketitle

\thispagestyle{fancy}


\section{THE SNR PARADIGM}
Even before the launch of the Fermi satellite, several supernova remnants (SNRs) have been detected in the $\gamma$-rays by Cherenkov telescopes, attesting to the presence of multi-TeV particles at their blast waves.
This evidence supports the \emph{SNR paradigm} for the origin of Galactic cosmic rays (CRs), which relies on the fact that: i) SNRs can account for the proper energetics \citep{baade-zwicky34}; ii) non-relativistic, strong shocks are expected to accelerate particles distributed in energy as an universal power-law, $\propto E^{-2}$ \cite[e.g.][]{bell78a,blandford-ostriker78}. 
In SNRs signatures of accelerated electrons have been found since the '50s, in the shape of radio synchrotron emission; however, the presence of relativistic nuclei can be directly probed only in the $\gamma$-rays resulting from the decay of $\pi_0$s produced in nuclear collisions between accelerated ions and the background plasma

In the TeV band, the $\gamma$-ray spectrum almost invariably shows a high-energy cut-off, which  reflects the cut-off of the distribution of accelerated particles.
With only this information available, it is impossible to disentangle whether the emission is \emph{hadronic}, i.e., due to neutral pion decay, or \emph{leptonic}, i.e., due to the inverse-Compton scattering of relativistic electrons on some photon background.

The biggest difference between the two scenarios above is the \emph{slope} of the photon energy distribution: while pion decay produces a spectrum parallel to the proton one, inverse-Compton produces a significantly flatter spectrum. 
Since acceleration is rigidity-dependent, at energies where synchrotron losses are negligible, electron and ion spectra are parallel ($\propto E^{-q}$). 
As a consequence, the expected photon spectrum is $\propto\nu^{-(q+1)/2}$ ($\propto\nu^{-q}$) in the leptonic (hadronic) scenario.
The interesting range of photon energies that allows us to test the very nature of the $\gamma$-ray origin falls exactly in the Fermi-LAT band.

\section{EVIDENCE OF STEEP SPECTRA}
\subsection{$\gamma$-ray emission from SNRs}
\begin{figure}[b]
\centering
\includegraphics[trim=40px 10px 10px 28px, clip=true, width=0.48\textwidth]{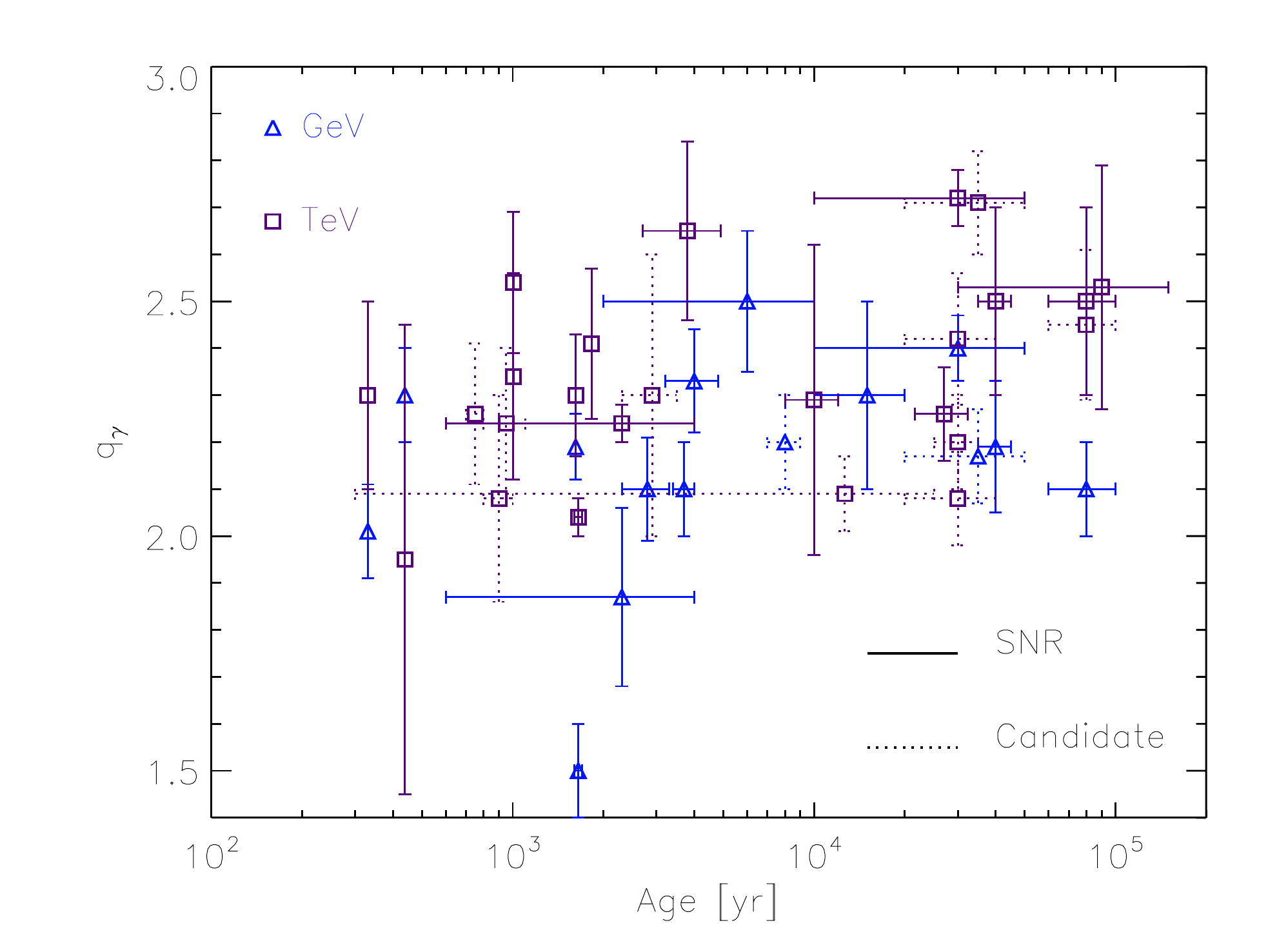}
\caption{Spectral slopes, $q_{\gamma}$, of $\gamma$-ray--bright SNRs. From table 1 in \cite{gamma} (see references therein), updated with data of W41, Puppis A and Cygnus Loop.} \label{gammaall}
\end{figure}

From a collection of the spectral indexes of SNRs detected in the GeV and/or in the TeV band \citep{gamma,mb13}, it is rather clear that the leptonic scenario is often disfavored: except RX J1713.7-3946 (and, possibly, Vela Jr.), all of the sources show spectra steeper than $E^{-2}$.
Quite interestingly, when a SNR is detected both in the GeV and in the TeV band, its spectrum can usually be fitted with a single power-law, suggesting a coherent origin for the $\gamma$-ray emission.
The alternative is to fine-tune the leptonic contributions of inverse-Compton and relativistic bremsstrahlung, as it has been proposed, e.g., for Cas A \citep{CasAFermi}. 

The photon spectral slope typically reads $q_\gamma=2.2-2.5$; 
such $\gamma$-ray spectra, to be explained as due to inverse-Compton, would require an electron distribution as steep as $E^{-3.5}-E^{-4}$, at odds with radio/X-ray observations, and with any rigidity-dependent acceleration mechanism at strong shocks. 

\subsection{\label{sec:MCs} SNRs and molecular clouds}
Some $\gamma$-ray--bright SNRs, however, may belong to a distinct class of sources because of their interaction with molecular clouds (MCs) \citep[see, e.g.,][]{cs10}. 
Some SNRs (W44, W49B, W28N and IC 443, for instance), show a peculiar cut-off around 3--10 GeV,  possibly related to the interaction of escaping CRs with MCs as well.
The correlation of $\gamma$-rays with high gas densities is a strong hint that the emission is of hadronic nature. 
Very recently \cite{pizero} confirmed this hypothesis by detecting the characteristic pion-decay feature in IC 443 and W44 \citep[see also][for AGILE measurements]{w44AGILE}.
Nevertheless, one must bear in mind that 1) MCs may shine also because of diffuse Galactic CRs \citep[e.g][]{galridge}; 2) modeling the expected spectrum from a MC is not straightforward, because of the poorly-understood diffusion of CRs close to sources and in very dense environments \citep{gab07}.
SNRs close to, or interacting with, MCs show hadronic emission, but might not be probing freshly accelerated particles.
Certainly, they are very hard to model in a self-consistent fashion, also because of the lack of emission in other bands of the spectrum. 

\subsection{Anisotropy of diffuse Galactic CRs}
There is another observational evidence that the spectrum injected in the Galaxy by SNRs must be steeper than $E^{-2}$.
The primary to secondary ratios and the anisotropy in the CR fluxes observed at Earth can be accounted for only if the residence time in the Milky Way has a mild dependence on the CR rigidity: $\propto (E/Z)^{-\delta}$, with $\delta\sim 1/3$ \cite[e.g.][]{vladimirov+11,ab12b}.
Since $\delta$ represents the correction to the injection spectrum due to propagation in the Galaxy, the constraint $q+\delta\approx 2.75$ implies that SNRs inject CRs with spectra steeper than $E^{-2}$.
Differential escape of CRs can steepen the instantaneous spectra probed in the $\gamma$-rays by  $0.1-0.2$ in slope \citep{crspectrum,nuclei}, but the most straightforward scenario points towards particles being accelerated with spectra $\propto E^{-2.3}-E^{-2.4}$.

\section{THEORY VS OBSERVATIONS}
Steep spectra are also a challenge to the current theory of diffusive shock acceleration (DSA).
The general consensus among non-linear models of DSA, namely models that account for the dynamical back-reaction of CR pressure and energy, has traditionally been that the more efficient the particle acceleration, the flatter the CR spectrum \citep[see][for a review]{malkov-drury01}. 
$E^{-2}$ is the steepest spectrum expected in standard non-linear DSA, and it corresponds to a very inefficient particle acceleration (less than 1\%), at odds with the requirement of the SNR paradigm (10-30\%).
This is not a small correction, but a systematic discrepancy: if acceleration is efficient, the canonical theory predicts the CR spectrum to be invariably \emph{flatter} than in the test-particle limit. 
In the next section we outline a possible way of reconciling the theory of efficient DSA with the steep spectra observed in SNRs. 

\subsection{Magnetic field amplification}
\begin{figure}
\centering
\includegraphics[trim=20px 0px 20px 10px, clip=true, width=0.48\textwidth]{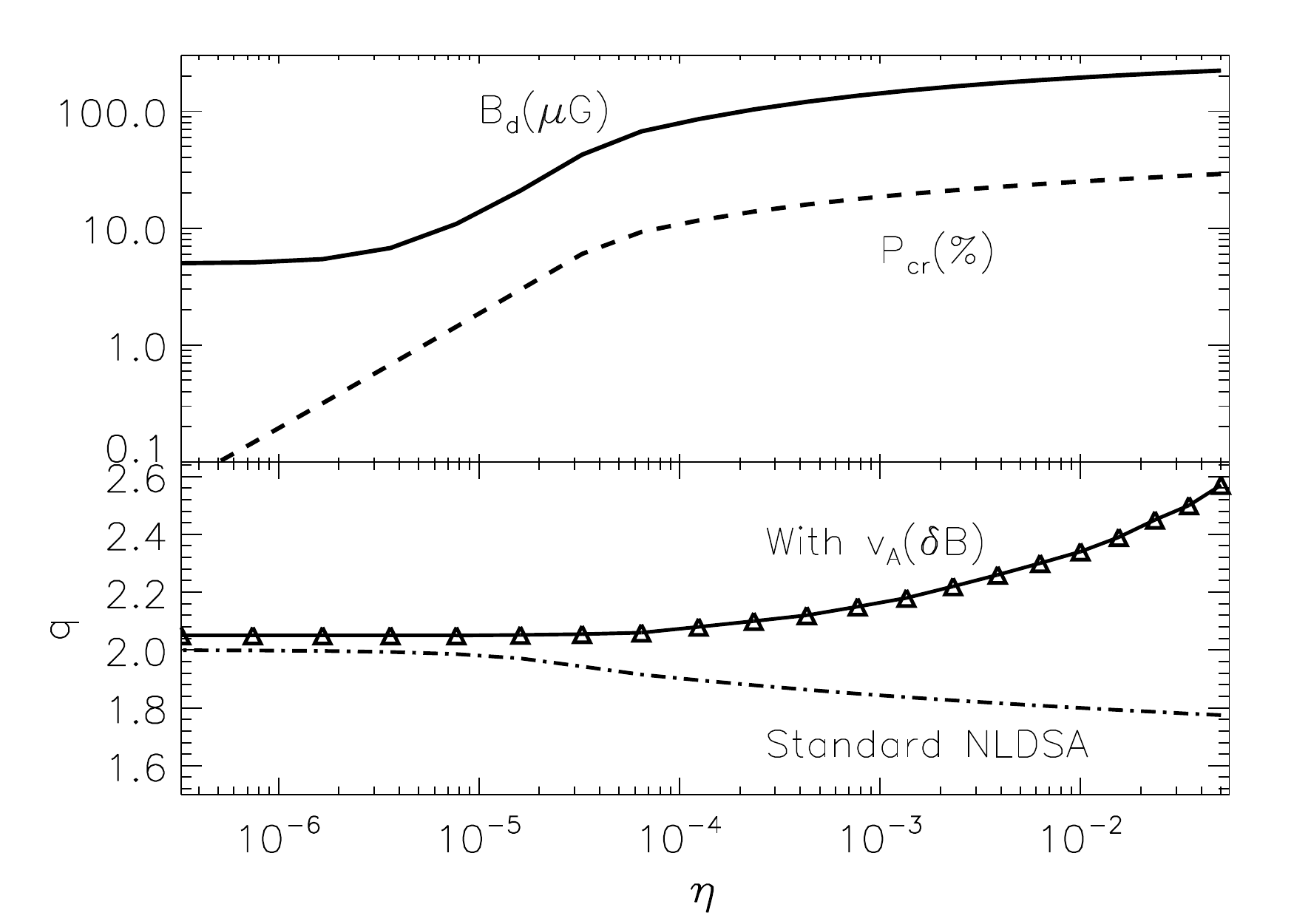}
\caption{\emph{Top panel}: downstream magnetic field, $B_d$, and CR acceleration efficiency, $P_{cr}$, as a function of the fraction of injected particles, $\eta$. 
\emph{Bottom panel}:  slope $q$ obtained with the scattering center velocity calculated in the amplified  $v_A(\delta B)$, or in the background field (Standard Non-Linear DSA). 
The values are for a SNR in the early Sedov stage \citep[about 3000yr, from][]{efficiency}.\\
} \label{eff}
\end{figure}
SNR blast waves are associated with enhancements of the magnetic field by factors of $10-100$ with respect to the mean Galactic value, as it follows from several observational facts:
1) young SNRs show non-thermal X-ray rims, whose thickness can be explained as the synchrotron loss-length of multi-TeV electrons radiating in magnetic fields as large as $100-500\mu$G  \citep[e.g.,][]{V+05,P+06};
2) X-ray variability on a time-scale of a few years is consistent with ---localized--- mG magnetic fields \citep{uchiyama+07}; 
3) fitting radio-to-X-ray data typically shows electrons to be synchrotron-cooled above a critical energy, at which the loss-time equals the age of the remnant. 
In Tycho, for instance, this constraint returns an average post-shock field of about 200$\mu$G \citep{tycho};
4) the lack of detection, within Chandra's resolution, of X-rays ahead of the shock of SN1006  suggests that the field is amplified in a CR-induced precursor \citep{gio-sn1006}.
Magnetic field amplification is also important to effectively scatter particles back and forth across the shock, potentially allowing SNRs to accelerate protons up to a few PeV, and heavier ions to larger energies, proportionally to their charge.  
Self-generated fields may even be dynamically relevant, in that the magnetic pressure may become comparable to, or even larger than, the plasma pressure, hence reducing the pre-shock magnetosonic Mach number \citep{jumpl}.

The most intriguing aspect of magnetic field amplification, though, is that these strong fields are expected to be generated by plasma instabilities driven by the same accelerated particles \citep[see][for a recent review]{ber12}.
If the growth rate is large enough to allow the turbulence to become nonlinear, as required to amplify the field, the local Alfv\'en velocity may increase significantly. 
If we take the Alfv\'en velocity $v_A(\delta B)$ as a proxy for the velocity of the magnetic irregularities particles scatter against, the CR diffusion-convection equation has to be accordingly modified, eventually leading to a modification of the CR spectrum.
The very reason behind this phenomenon is that the CR spectral slope is actually determined by the velocity jump $r=v_u/v_d$ felt by particles diffusing between upstream and downstream (or viceversa).
For strong gaseous shocks, $r=4$ and $q=\frac{r+2}{r-1}=2$; if relativistic CRs are present, $r>4$ and $q<2$. 
However, if $v_A$ is not negligible with respect to $v_u$, then we may have $r=\frac{v_u-v_A}{v_d}<4$ and $q>2$, at least in the simple case of excitation of resonant Alf\'enic modes \citep[e.g.][]{bell78a}, where self-generated waves upstream propagate against the fluid (i.e., in the direction opposite to the CR gradient), and downstream are almost isotropic  ($v_{A,d}\approx 0$).

If acceleration is efficient, the CR pressure tends to make the spectrum flatter, while field amplification may provide an opposite backreaction that steepens the spectrum.
In order to see which effect dominates over the other, a self-consistent model is required.
Using a semi-analytical approach to non-linear DSA, it is possible to show that the more efficient the CR acceleration, the more effective the magnetic field amplification, and the steeper the spectra of the accelerated particles \citep{efficiency}.
Such a trend is showed in Fig.~\ref{eff}, where CR pressure, downstream magnetic field, and spectral slope are plotted against the fraction of particles injected into the acceleration mechanism, $\eta$, for a SNR in its early Sedov stage \citep[see][for more details]{efficiency}. 
For $\eta\geq 10^{-4}$ the acceleration efficiency $P_{cr}$ (expressed as the pressure in CRs over the bulk pressure) saturates to $10-20$\%, but the spectrum becomes steeper and steeper. 
The non-linear feedback of the amplified fields may account, at the same time, for efficient acceleration, for the amplified magnetic fields (a few hundred $\mu$G), and for the spectral slopes inferred from $\gamma$-ray observations ($q=2.1-2.5$).
   
\subsection{Tycho's SNR}   
\begin{figure}
\begin{center}
\includegraphics[trim=0px 0px 0px 0px, clip=true, width=0.45\textwidth]{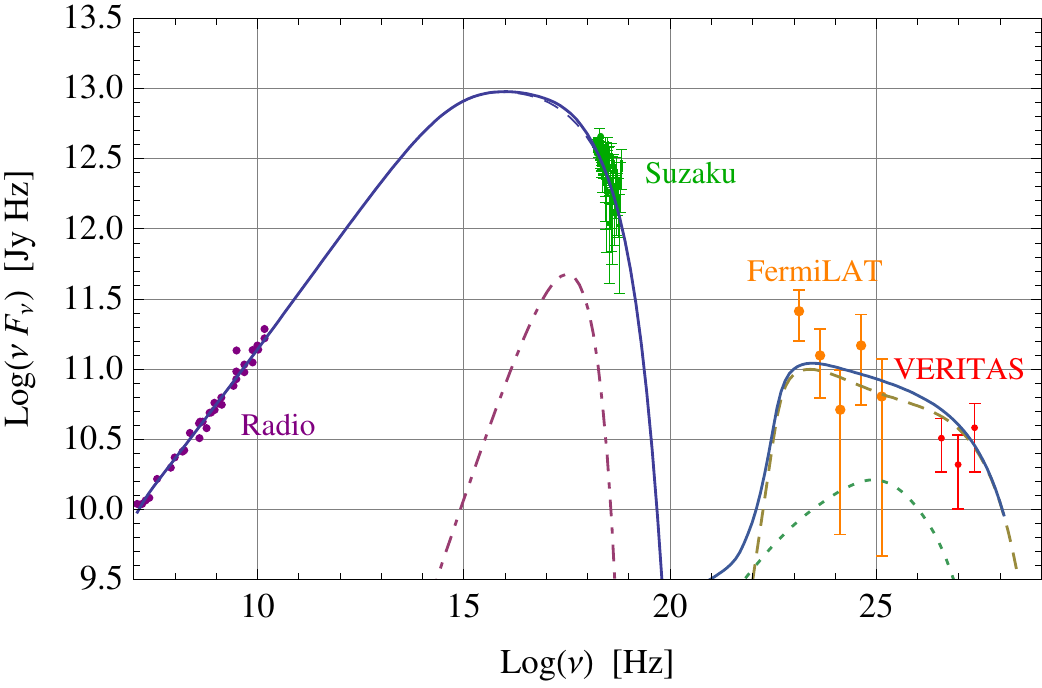}
\caption{Tycho's broadband emission.
Curves show the calculated synchrotron, inverse Compton, bremsstrahlung and
pion decay emission.
Refer to \cite{tycho} for the experimental data and for more details. }
\label{fig:multiwave}
\end{center}
\end{figure}

Tycho is probably the best SNR where to study DSA: it corresponds to the historical SN1572 (so that its age is known), and it is the remnant of a type-Ia explosion, which puts a constraint on the SN energy ($\approx 10^{51}$erg) and on the ejecta mass (about one solar mass); 
both kinematic and echo-light measurements set its distance within 3 and 4 kpc, implying a circumstellar density of $\sim0.3$ protons cm$^{-3}$.
Also, detailed maps in radio and X-rays are available, and the shell presents a rather clean spherical symmetry. 
All of this information significantly reduces the number of free parameters, making modeling Tycho's broadband emission an over-constrained problem.
There are two parameters that need to be tuned: the fraction of protons injected into the acceleration mechanism, and the electron-to-proton ratio at relativistic energies.
The maximum proton energy can be estimated by equating the acceleration time with the age of the system, which is also consistent with requiring the diffusion length of the highest-energy particles to be about one tenth of the SNR radius.
Given a reliable analytical description of the SNR evolution \citep{TMK99}, a self-consistent treatment of the shock dynamics including CR and magnetic contributions, and a model for magnetic field amplification, it is possible to work out the expected multi-wavelength, radially-resolved, map of the non-thermal emission with two free parameters only \citep{tycho}.

The less solid parts of the calculation are the estimate of the growth rate of the resonant streaming instability, extrapolated from the quasi-linear regime, and the assumption that the Alfv\'en velocity traces the velocity of the scattering centers.
Also, we assume Bohm diffusion, in the amplified field.
Strictly speaking, the Bohm limit corresponds to resonant scattering onto self-generated Alfv\'en waves of amplitude $\delta B/B\sim 1$, and it is achieved when the CR energy distribution is $\propto E^{-2}$.

Recently, \cite{tycho-bv} criticized the adoption of a Bohm-like diffusion coefficient when the CR spectrum is steeper than $E^{-2}$, arguing that in this case diffusion should be less effective, thereby reducing the maximum energy achievable at any given time.
Firstly, the proton spectrum we obtained retains a hint of the curvature typical of CR-modified shocks, flattening at the highest energies, as pointed out in fig.~4 of \cite{tycho}. 
Secondly, and most importantly, when $\delta B/B\gg 1$ Bohm diffusion is not more than a heuristic assumption:
the normalization of the diffusion coefficient, adopted by \cite{tycho-bv} as well, must be a factor of $\delta B/B\sim 30-50$ smaller than the one predicted by the quasi-linear theory to allow CR acceleration up to a few hundred TeV.
In this respect, a deviation of less than 10\% in the energy dependence of the diffusion coefficient seems indeed acceptable. 
The main criticism than can be raised against \emph{all} of the models proposed is that the magnetic field must be amplified even far upstream, on the diffusion scale of the highest-energy particles. 
Resonant streaming instability is unlikely to be relevant there, and magnetic field amplification has probably to rely on instabilities triggered by escaping particles, like the non-resonant hybrid instability \citep{bell04}.
If this is the case, however, it is still not clear whether the excited modes, which have short-wavelengths and are non-resonant with ions in terms of polarization, can scatter CRs effectively. 

A crucial difference between our model and the one proposed by \cite{tycho-bv} is that in ours the magnetic field is calculated according to a simple but physically-motivated recipe, and not put by hand.
Therefore, the spectral steepening induced by the enhanced velocity of the scattering centers is not arbitrary, but it is consistently related to the resulting CR acceleration efficiency and to the inferred magnetic fields.
Conversely, \cite{tycho-bv} explain Tycho's steep $\gamma$-ray spectrum by introducing an additional contribution of CR protons accelerated at several low-Mach--number reverse shocks produced by the interaction of the blast wave with small, dense clouds filling the circumstellar medium.
This scenario, however, requires quite a fine-tuning of the properties of these weak shocks in terms of densities, temperatures, velocities, magnetic fields, and electron synchrotron emissivity to match the $\gamma$-ray emission without violating any other observable, which the authors fit with the "canonical" CR population only \citep{tycho-bv-old}. 
It is quite unlikely for such a conspiracy to work also in many other SNRs, while steep spectra are ubiquitous (Fig.~\ref{gammaall}).
In this respect, it seems more natural to conceive the hypothesis that CRs may be accelerated with spectra steeper than those predicted by the standard non-linear DSA theory.
   
\section{HYBRID SIMULATIONS OF COLLISIONLESS SHOCKS}    
\begin{figure}
\begin{center}
\includegraphics[width=0.49\textwidth]{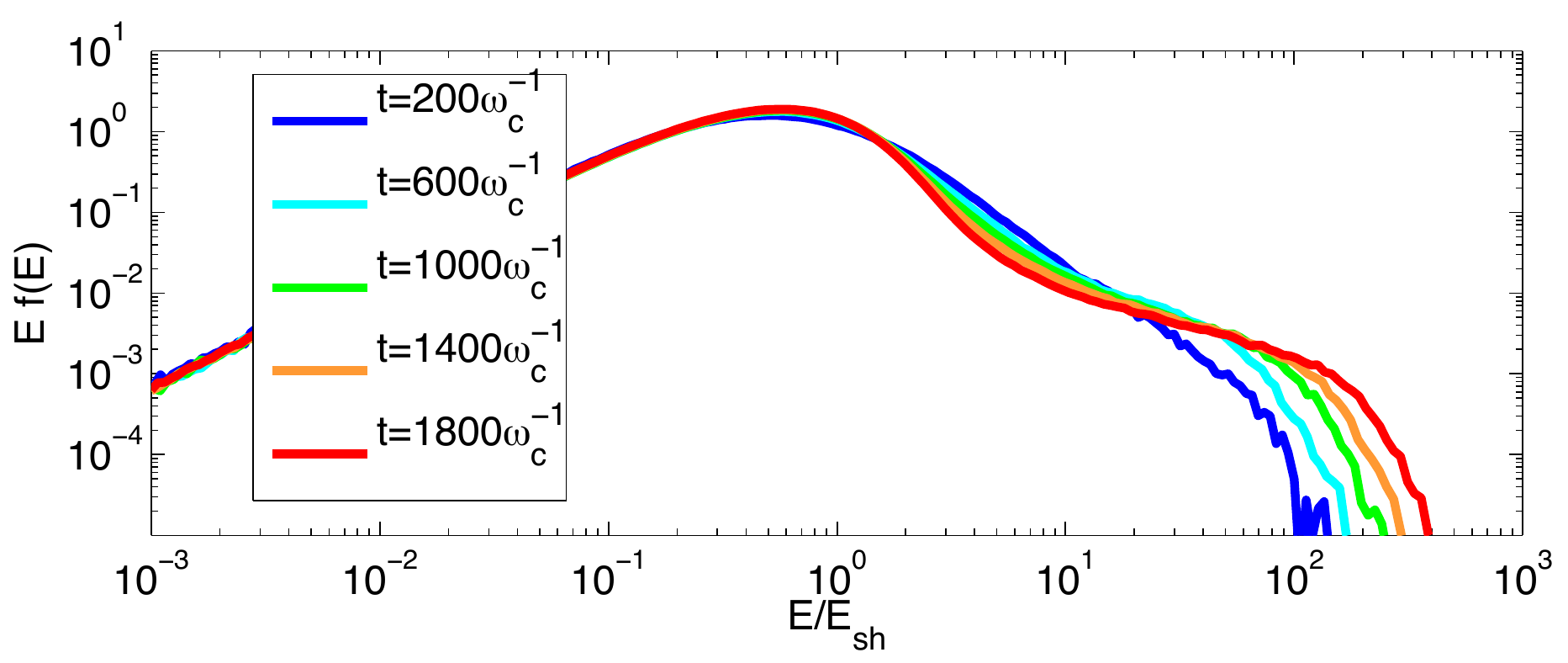}
\caption{Evolution of the post-shock energy spectrum in a hybrid simulation of a parallel $M=20$ shock. 
Time is units of (ion cyclotron frequency)$^{-1}$, and energy is in $E_{sh}=\frac{m}{2}V_{sh}^2$, with $V_{sh}$ the shock velocity. 
The spectral slope matches the DSA prediction for strong shocks.}
\label{fig:evo}
\end{center}
\end{figure}

As it is clear also from the discussion above, any theoretical model aiming to account for the many  observational constraints requires a detailed understanding of the plasma instabilities triggered by the accelerated particles, including the nature of the fastest-growing modes, their saturation, and their effects in scattering CRs. 
Also, particle injection has not been understood from first-principles, yet, and any non-linear approach to shock acceleration has to model this crucial ingredient through some phenomenological recipe. 
Finally, we need to prove that Fermi acceleration may be as efficient as inferred, and that it can sustain itself by generating enough magnetic turbulence to allow the maximum energy to increase with time. 

These crucial pieces of physics involve very small plasma scales, and must be tackled with numerical approaches. 
Particle-in-cell techniques can capture the intrinsically nonlinear nature of the interplay between accelerated particles and the electromagnetic field: single-particle trajectories are calculated according to the Lorentz force produced by self-consistent currents.
Nevertheless, PIC simulations are computationally very expensive.
To mitigate this limitation, the so-called \emph{hybrid} limit can be used: (massless) electrons are taken as a neutralizing fluid, while ions are treated kinetically \citep[see][for a review]{lipatov02};
this approach still captures both the shock dynamics and the ion anisotropy, which drives the plasma instabilities relevant for magnetic field amplification.

\begin{figure}
\includegraphics[trim=40px 50px 40px 55px, clip=true, width=.45\textwidth]{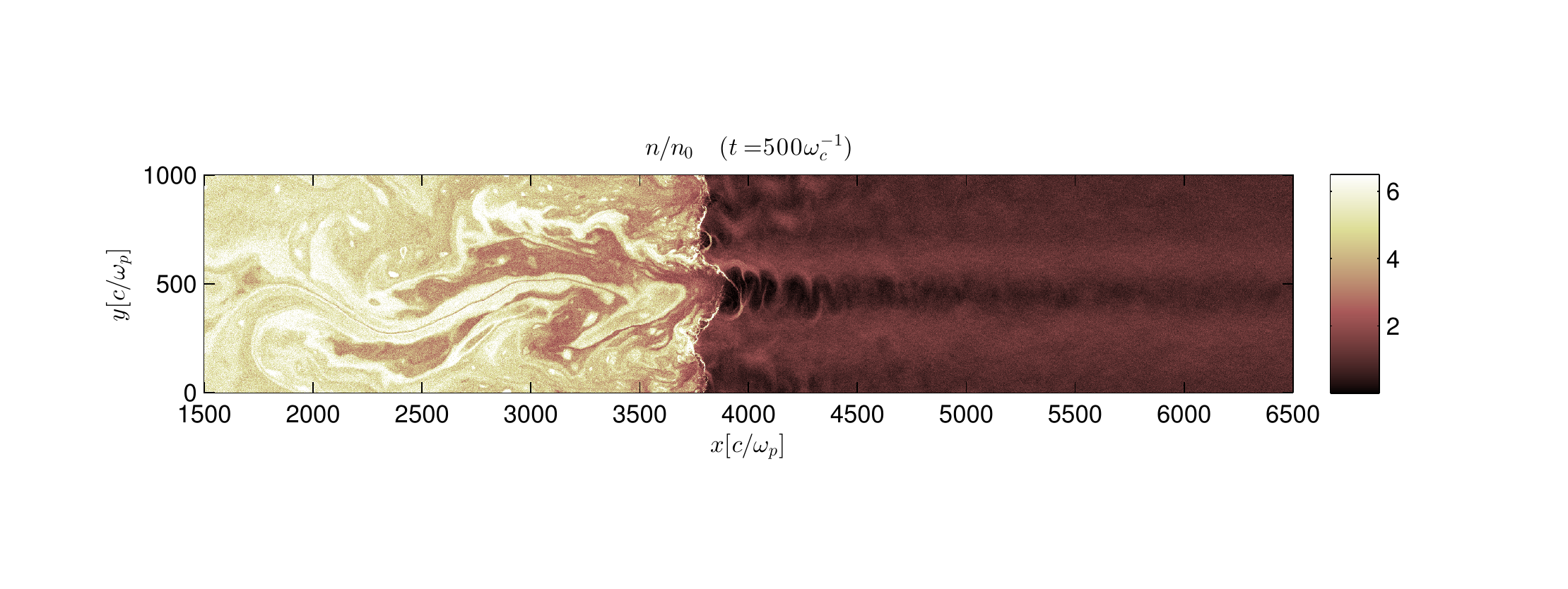}
\includegraphics[trim=40px 50px 40px 50px, clip=true, width=.45\textwidth]{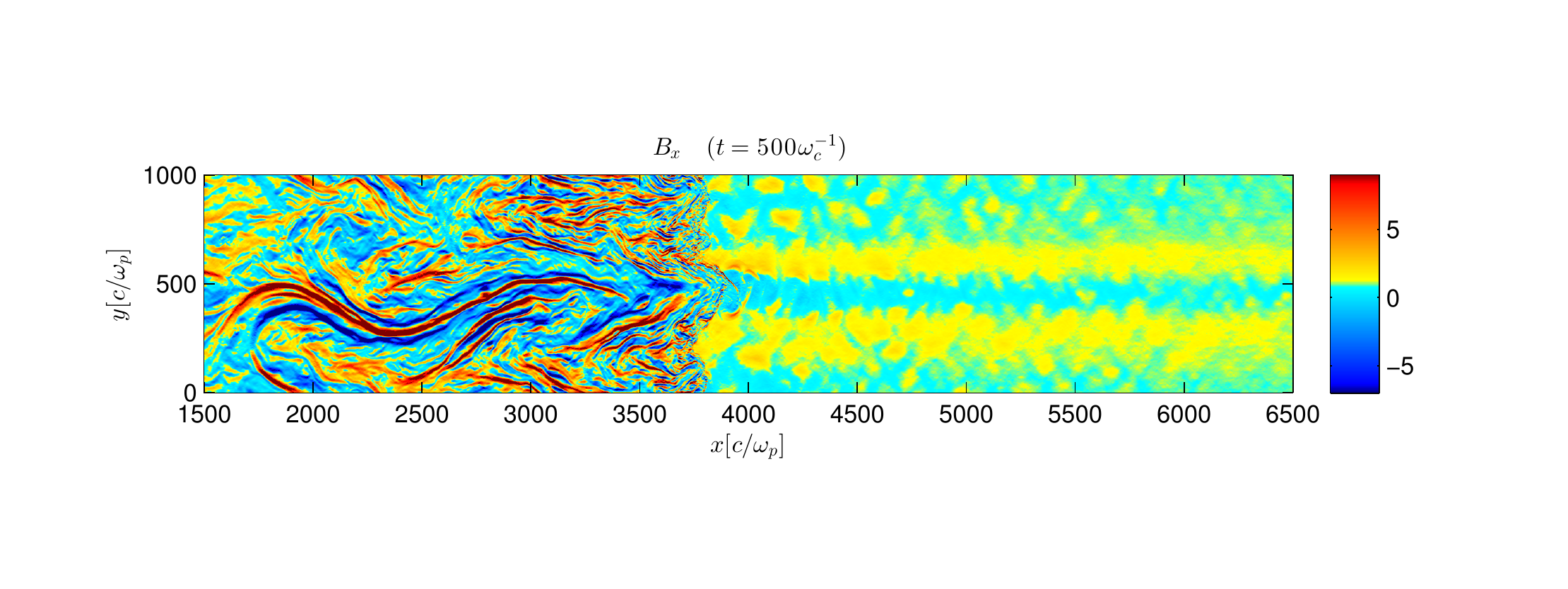}
\includegraphics[trim=40px 50px 40px 50px, clip=true, width=.45\textwidth]{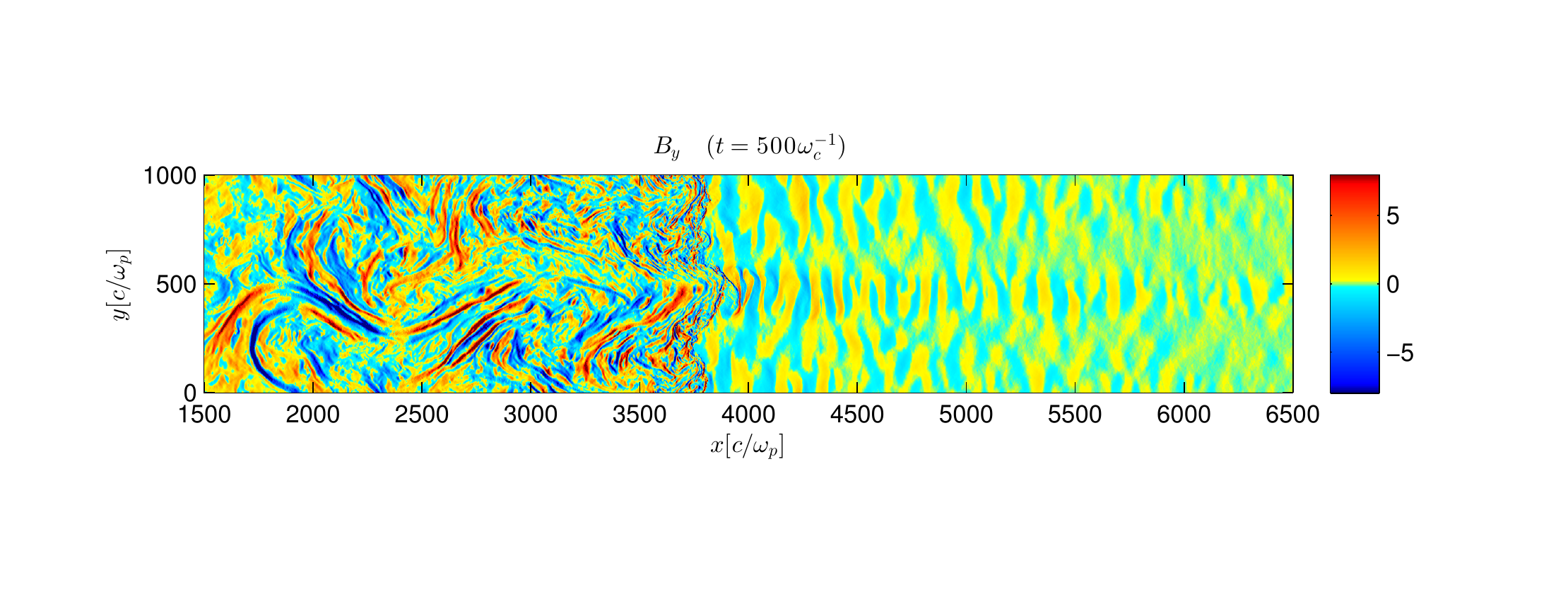}
\includegraphics[trim=40px 50px 40px 50px, clip=true, width=.45\textwidth]{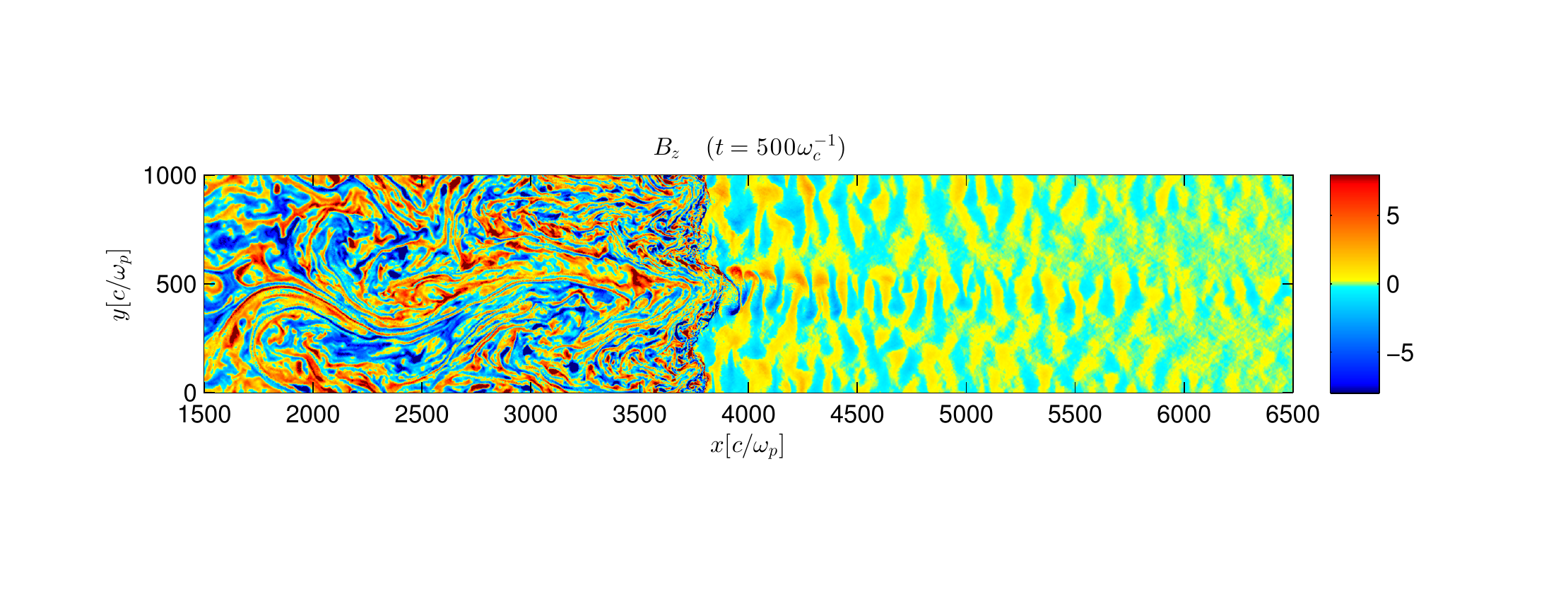}
\includegraphics[trim=40px 50px 40px 50px, clip=true, width=.45\textwidth]{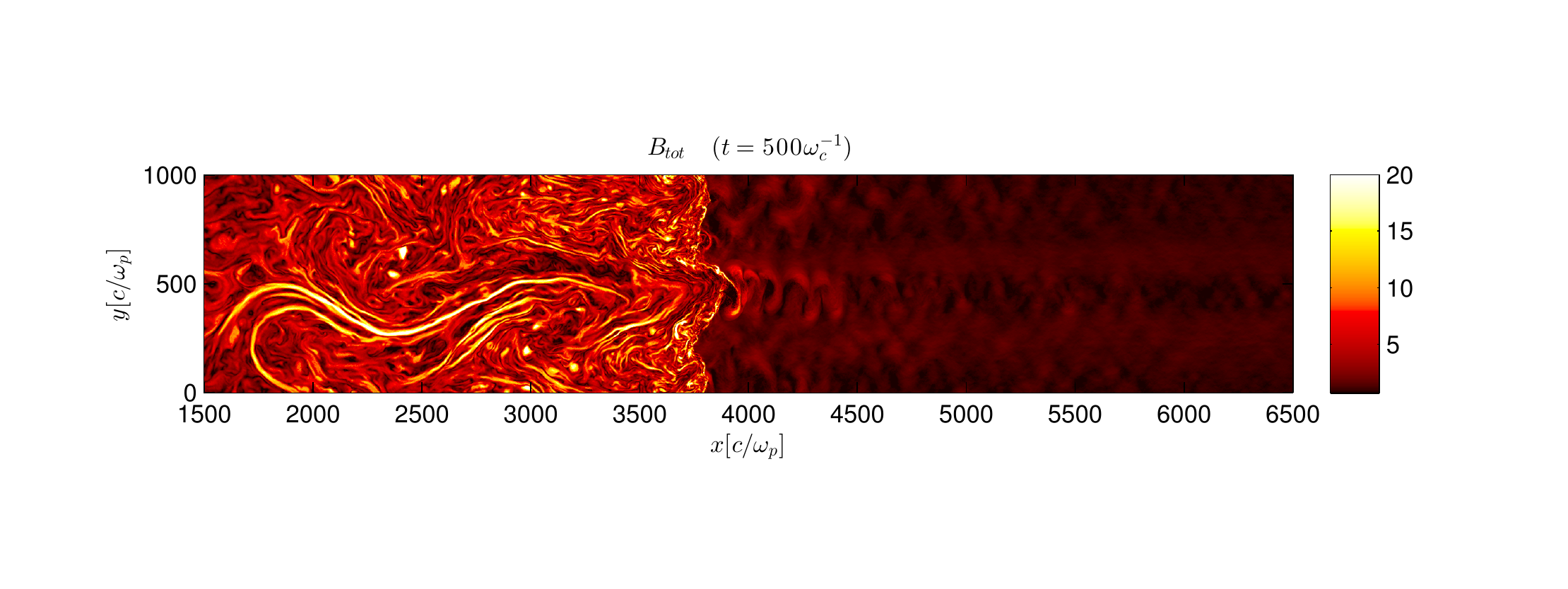}
\includegraphics[trim=40px 50px 40px 50px, clip=true, width=.45\textwidth]{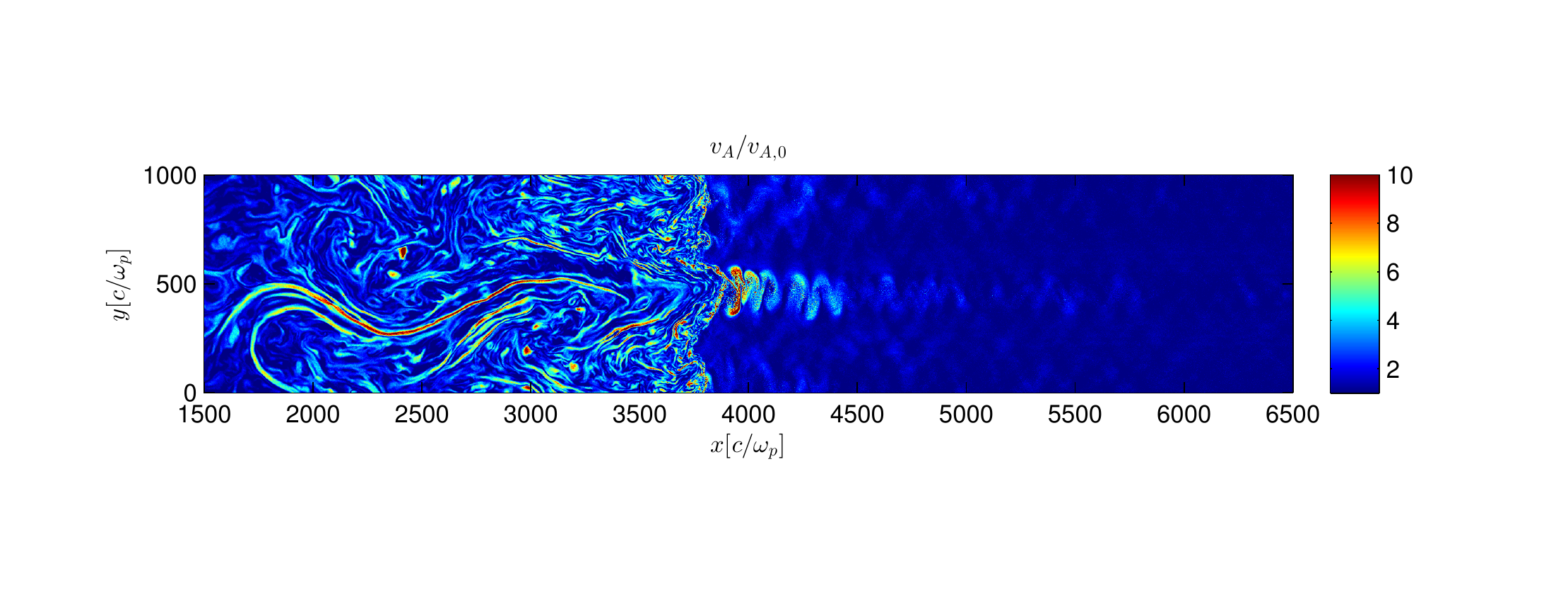}
\caption{\label{fig:par}
Density, parallel ($B_x$), transverse ($B_y$,$B_z$) and total ($B_{tot}$) magnetic field, and Alfv\'en velocity $v_A$ for a 2D simulation of a $M=30$ parallel shock, at $t=500\omega_c^{-1}$. 
Quantities are normalized to their unperturbed values.
A filament with $B_{tot}\approx 20B_0$ and knots with $B_{tot}\approx 40B_0$ are clearly visible \citep{filaments}.}
\end{figure}

Here we summarize some results recently obtained with the newtonian (i.e., non-relativistic) code \emph{dHybrid} \citep{gargate+07}. 
We consider a parallel shock (background field $B_0$ parallel to the shock velocity $V_{sh}$) with equal sonic and Alfv\'enic Mach numbers, $M\simeq 20$.
The computational box measures $10^5\times10^2$ in units of ion skin depths $c/\omega_p$, with $\omega_p$ the ion plasma frequency, and 2 cells per ion skin depth are used;
despite the box being 2D, three components of particle velocities and electromagnetic fields are accounted for. 

Fig.~\ref{fig:evo} shows the evolution of the post-shock ion energy distribution: the spectrum develops a non-thermal power-law tail, whose extension increases with time. 
Very interestingly, the spectral slope of the accelerated particles is the one predicted by Fermi acceleration at strong shocks, namely $f(p)\propto p^{-4}$ in momentum space, which translates into $E^{-1.5}$ for non-relativistic particles ($p^2=2mE$). 
If the spectrum extended into relativistic energies, $p\propto E$, and the standard $f(E)\propto E^{-2}$ would be recovered. 
The total energy in the non-thermal tail is about 10\% of the shock bulk energy, attesting to the ability of parallel shocks to efficiently accelerate ions [Caprioli and Spitkovsky, in preparation].  
Kinetic simulations do not allow to test all of the nonlinear effects mentioned above, yet, but they allow us to self-consistently study the development of the instabilities induced by energetic particles. 
Unprecedentedly-large 2D and 3D hybrid simulations \citep{filaments} showed that high-Mach number parallel shocks are prone to the CR-driven \emph{filamentation instability}:
the streaming of accelerated particles upstream of the shock produces cavities of rarefied plasma and low-magnetic field, surrounded by a net of dense filaments with strong fields \citep[see also][]{rb12}.
The cavitation of the upstream, on scales comparable with the gyroradii of the accelerated ions, produces a significant increase of the magnetic field and of the local Alfv\'en velocity (see Fig.~\ref{fig:par}); this effect may reveal itself crucial for enhancing particle diffusion.
Moreover, the advection of these cavities through the shock produces bubbles subject to (impulsive) Rayleigh--Taylor instability, which triggers turbulent downstream motions enhancing the magnetic field amplification. 
Some knots in Fig.~\ref{fig:par} carry a magnetic field $\sim 40$ times larger than $B_0$, and preliminary runs at $M=50$ show that the maximum field can easily become as large as $\sim 100 B_0$. 
This localized strong fields may be consistent, for instance, with the X-ray variability reported by \cite{uchiyama+07}.   
The downstream bubbles are filled by plumes of hot plasma, which stretch the magnetic field along the shock direction of propagation (Fig.~\ref{fig:par}: density, $B_{x}$, and $B_{tot}$ plots).
These coherent, elongated structures, which are predicted to develop only where the shock is parallel, may have an observational counterpart in the pattern of radial X-ray stripes detected in Tycho \citep{eriksen+11}.

\section{CONCLUSIONS}
More than a century after the discovery of CRs, and many decades after the discovery that electrons are accelerated in SNRs, we are finally getting direct evidence of the acceleration of ions in these environments.
GeV observations by Fermi-LAT are complementing TeV data obtained with Cherenkov telescopes, opening an unprecedented window on the non-thermal emission of SNRs.
In some cases the $\gamma$-rays observed are unequivocally of hadronic origin \citep{pizero}, but extracting quantitative constraints on the mechanisms responsible for ion acceleration requires a non-trivial theoretical modeling.
When the emission comes from MCs illuminated by nearby SNRs, there is a degeneracy between the injection spectrum and the correction due to the poorly-known diffusion around the sources. 
Also, when shocks are directly interacting with dense, partially-neutral gas, the dynamics is strongly perturbed by charge-exchange processes \citep{neutri}.
Modeling these shocks requires a kinetic description of the neutral distribution, but the effort is rewarded with an additional diagnostics, based on the study of H$\alpha$ emission, of ion temperature and CR acceleration efficiency \citep{GLR07,raymond+11,neutri3}. 

The most straightforward way to probe CR acceleration at SNRs is to look at SNRs non interacting with MCs \citep[which include Tycho: see][]{tl11}, whose steep $\gamma$-ray spectrum challenges the current models of efficient particle acceleration at shocks.
A crucial open problem is to understand in greater detail the mechanisms that amplify the magnetic field up to the inferred values, mechanisms that may be important also for regulating the spectral slopes of accelerated particles.
Kinetic simulations play a pivotal role in investigating the complex particle-field interplay from first-principles.
At the core of the acceleration process, however, always lies the pioneering idea envisioned by Enrico Fermi, and namely that a particle can increase its energy by being repeatedly reflected in head-on collisions \citep{Fermi49}.
More than 60 years later, a satellite named after him is providing us with very strong evidence that SNRs are indeed the sources of Galactic CRs.

\bigskip 
\begin{acknowledgments}
The author wishes to thank A. Spitkovsky and G. Morlino for their collaboration, and L. Gargat\'e for his help with \emph{dHybrid}.
This research was supported by NSF grant AST-0807381 and NASA grants NNX09AT95G and NNX10A039G. Simulations were performed on the computational resources supported by the PICSciE-OIT TIGRESS High Performance Computing Center and Visualization Laboratory. This research also used the resources of the National Energy Research Scientific Computing Center, which is supported by the Office of Science of the U.S. Department of Energy under Contract No. DE-AC02-05CH11231, and Teragrid/XSEDE's Ranger under contract No.\ TG-AST100035.
\end{acknowledgments}

\bibliography{fermi12}
\bibliographystyle{apsrev}
\end{document}